\documentclass[twocolumn,nofootinbib,preprintnumbers,amsmath,amssymb]{revtex4}

\usepackage{graphicx}% Include figure files
\usepackage{dcolumn}% Align table columns on decimal point
\usepackage{bm}% bold math
\usepackage{amsmath}

\def \be {\begin{equation}}
\def \ee {\end{equation}}

\begin{document}

\title{No-Hair Theorem for Weak Pulsar}% Force line breaks with \\

\author{Andrei Gruzinov}

\affiliation{ CCPP, Physics Department, New York University, 4 Washington Place, New York, NY 10003
}

\begin{abstract}

  It is proposed that there exists a class of pulsars, called weak pulsars, for which the large-scale magnetosphere, and hence the gamma-ray emission, are independent of the detailed pattern of plasma production. The weak pulsar magnetosphere and its gamma-ray emission are uniquely determined by just three parameters: spin, dipole, and the spin-dipole angle. We calculate this supposedly unique pulsar magnetosphere in the axisymmetric case. The magnetosphere is found to be very close to (although interestingly not fully identical with) the magnetosphere we have previously calculated, explaining the phenomenological success of the old calculation.

  We offer only a highly tentative proof of this ``Pulsar No-Hair Theorem''. Our analytics, while convincing in its non-triviality, is incomplete, and counts only as a plausibility argument. Our numerics, while complete, is dubious.

  The plasma flow in the weak pulsar magnetosphere turns out to be even more intricate than what we have previously proposed: some particles, after being created near the star, move beyond the light cylinder and then return to the star.

\end{abstract}

\maketitle

\section{Introduction}

One would justifiably think that the pulsar magnetosphere and its emission are calculable\footnote{ The most important pulsar result of Fermi \cite{Fermi} is the estimate of the median pulsar efficiency -- about 15\%. This shows that the magnetosphere cannot be calculated without calculating the emission, except maybe in a few cases.}  only together with the plasma production. This is true for strong and dying pulsars, but (we propose) there exists a class of pulsars, called weak pulsars, whose magnetosphere and gamma-ray emission are independent of the precise plasma production mechanism. Weak pulsars form  a three-parameter family, the parameters being spin ($\Omega$), magnetic dipole ($\mu$), and the spin-dipole angle ($\theta$). For simplicity we treat only the axisymmetric case, $\theta =0$. We work in pulsar units,
\be
c=\Omega=\mu=1,
\ee
and use cylindrical coordinates $(r,\phi ,z)$.

\section{Weak Pulsar}

The pulsar is {\it weak} if it is not {\it strong} and not {\it dying}. The pulsar is {\it strong} if pair production rate near the light cylinder (the surface $r=1$) is substantial, meaning $\gtrsim 1$. The pulsar is {\it dying} if the plasma production rate near the star is not high enough, so that the proper electric field near the star is not fully screened.

\begin{figure}[bth]
  \centering
  \includegraphics[width=0.48\textwidth]{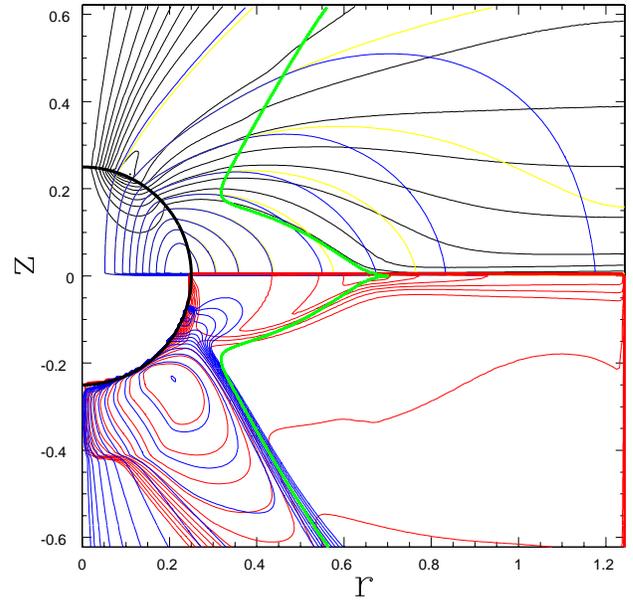}
\caption{Very thick green: the boundary between the force-free zone (where $E_0=0$) and the radiation zone. Upper half-plane: thin black: $A$, integer multiples of $0.1A_{\rm min}$, $A_{\rm min}=-0.36$;  thicker blue and thick yellow: $\phi$ and $\psi$, integer multiples of $0.1\psi _{\rm max}$, $\psi _{\rm max}=4.18$. Lower half-plane: thin blue: $(r^2+z^2)\rho _-$, thicker red: $(r^2+z^2)\rho _+$, isolines $0.003\times 2^k$, $k=1,2,...$} 
\end{figure}

\section{Numerics}

The calculated magnetosphere, Fig.1, is shown using the standard parametrization of the stationary axisymmetric electromagnetic field: 
\begin{align}
{\bf E} & =\left( -\partial _r\phi, 0, -\partial _z\phi \right) ,\\
{\bf B} & ={1\over r}\left( -\partial _z\psi, A, \partial _r\psi \right) .
\end{align} 
The calculation uses the full set of Aristotelian Electrodynamics (AE) equations:
\begin{align}
\dot{{\bf B}} & =-\nabla \times {\bf E},\\
\dot{{\bf E}} & =\nabla \times {\bf B}-{\bf j},
\\
{\bf j} & =\rho_+{\bf v}_+-\rho_-{\bf v}_-,
\\
\dot{\rho _\pm } & +\nabla \cdot (\rho _\pm {\bf v}_{\pm}) =\Gamma ,
\\
{\bf v}_{\pm} & ={{\bf E}\times {\bf B}\pm(B_0{\bf B}+E_0{\bf E})\over B^2+E_0^2},
\\
B_0^2 & -E_0^2 =B^2-E^2,~ B_0E_0={\bf B}\cdot {\bf E},~ E_0\geq 0.
\end{align}

In words: we solve Maxwell equations (4,5), with the current (6) from positrons and electrons of charge-normalized densities $\rho _\pm$ moving with velocities ${\bf v}_{\pm}$, and created at a rate $\Gamma$, (7). The velocities are given by the basic AE equation (8). Eq. (9) defines the proper electric field scalar and the proper magnetic field pseudoscalar. The basic AE equation and the gamma-ray emission accompanying this radiation-overdamped motion are fully derived in \cite{Gruzinov} (a) (page 2, first column, line 10). To the best of my knowledge, AE was first used by \cite{Finkbeiner}. 

All details, both physics (the plasma production rate $\Gamma$, the current {\bf j} inside the star, regularizations, ...) and numerics (grid, resolution, interpolations, ...) are given in the Appendix. 

\begin{figure}[bth]
  \centering
  \includegraphics[width=0.48\textwidth]{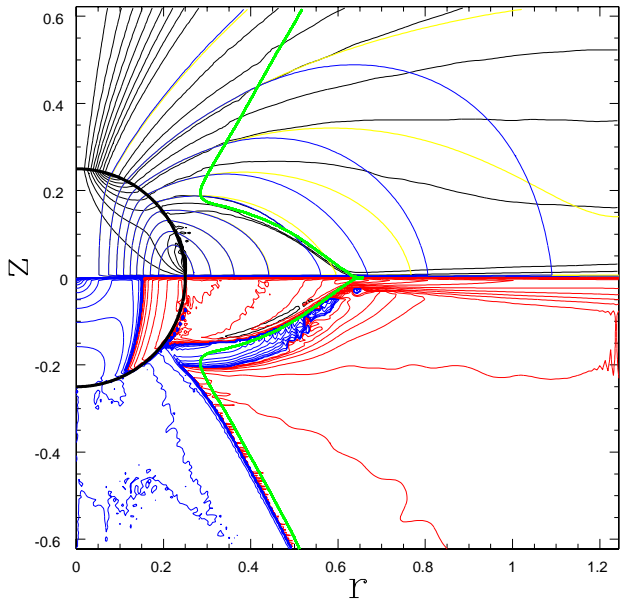}
\caption{Same as fig.1., with $A_{\rm min}=-0.33$, $\psi _{\rm max}=4.19$. In the lower half-plane: $(r^2+z^2)\rho $, thin blue where negative, thicker red where positive, isolines $\pm 0.003\times 2^k$, $k=1,2,...$}
\end{figure}

A very important result is that the full-AE magnetosphere (Fig.1) is quantitatively close to the magnetosphere obtained by a simplified AE calculation (Fig.2) which treats the plasma flow only implicitly, by adopting the following Ohm's law
\be
{\bf j}={\rho {\bf E}\times {\bf B}+|\rho |(B_0{\bf B}+E_0{\bf E})\over B^2+E_0^2}, ~~~~\rho \equiv \nabla \cdot {\bf E}
\ee
The agreement between the two simulations is ``very important'' for two reasons. First, we already know \cite{Gruzinov} (b) that the Ohm's-law calculation explains the weak pulsar phenomenology without a single adjustable parameter. Second, the fact that the two methods give very similar results is a strong argument in favor of the proposed uniqueness of the solution.

\begin{figure}[bth]
  \centering
  \includegraphics[width=0.48\textwidth]{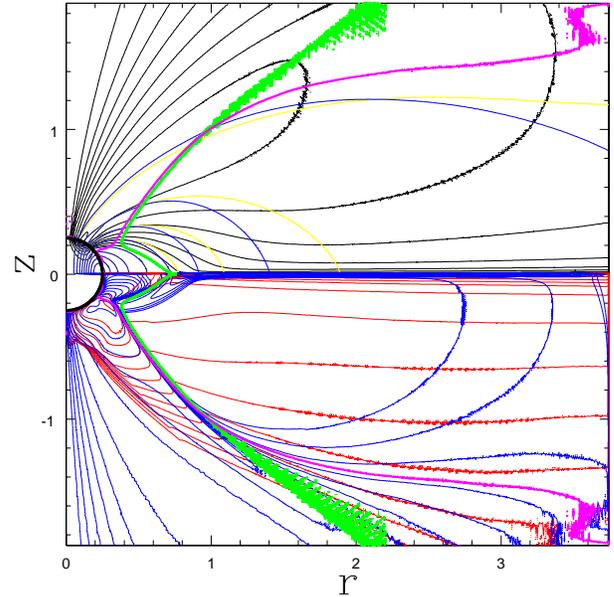}
\caption{Same as fig.1., with $A_{\rm min}=-0.30$, $\psi _{\rm max}=4.17$. Very thick magenta is the neutral surface -- the net charge is positive in the equator-containing domain bounded by the neutral surface and the stellar surface. } 
\end{figure}

\begin{figure}[bth]
  \centering
  \includegraphics[width=0.48\textwidth]{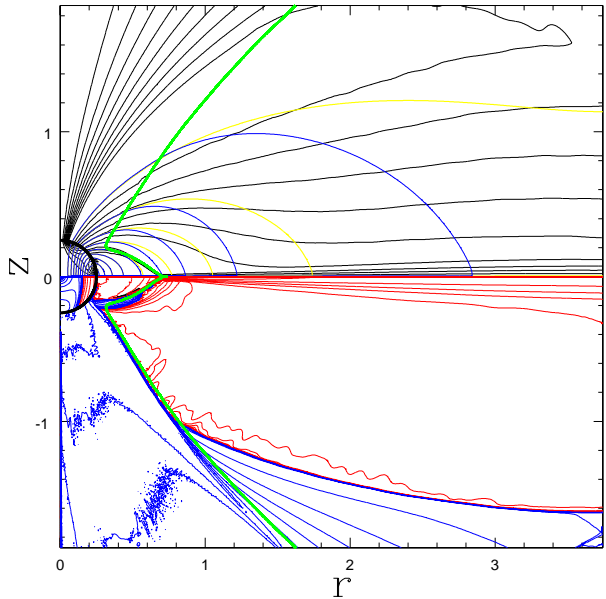}
\caption{Same as fig.2., with $A_{\rm min}=-0.26$, $\psi _{\rm max}=4.19$. }
\end{figure}

The reason for agreement between Fig.1 and Fig.2 must be as follows. From (8) we know that the true Ohm's law is 
\be\label{ohm1}
{\bf j}={\rho {\bf E}\times {\bf B}+P(B_0{\bf B}+E_0{\bf E})\over B^2+E_0^2},~~~~P=\rho_++\rho_-.
\ee
This coincides with the simplified Ohm's law (10) {\it iff} the plasma multiplicity is zero. As we see from Fig.1, the plasma multiplicity is zero only in the Radiation Zone, meaning that the simplified Ohm's, while exact in the Radiation Zone, is wrong in the Force-Free Zone. The agreement between Figg. 1 and 2 then means that the precise Ohm's law used in the Force-Free Zone is irrelevant. The very requirement that the Force-Free Zone stays force-free fixes the current in the Force-Free Zone.

Interestingly, and also important for the analytic arguments of the next section, when repeated in a larger box, the full-AE (Fig.3) and the Ohm's law calculations (Fig.4) do show slight differences. Now the plasma multiplicity in the Radiation Zone is non-zero, and the simple Ohm's law description (10) formally fails. However, as we see from Fig.4, the Ohm's law description remains very close to the full-AE result (Fig.3), and thus the phenomenological success of the Ohm's law calculation is not accidental. We will say more about the plasma flow in the next section.

\section{Analytics}

The most striking feature of the pulsar magnetosphere seen in Fig.3 is the co-existence of two well-defined domains: (i) the Force-Free Zone, where the proper electric field $E_0$ vanishes and (ii) the Radiation Zone, where the electromagnetic field has generic geometry with non-zero $E_0$. 

The very existence of the Force-Free Zone immediately raises two questions. First, for $E_0=0$, Eq.(9) leaves the sign of $B_0$ indefinite, meaning that one does not know how to calculate the velocities ${\bf v}_{\pm}$. Also, for $E_0=0$, the charges are no longer radiation-overdamped, and one cannot use AE at all. 

But our numerical experiments show that once the first problem is fixed by some regularization of formulas (9), the results become independent of the regularization. We propose that the requirement that there be a Force-Free Zone bounded by the Radiation Zone (where AE does apply) fully fixes the magnetosphere and the densities $\rho _\pm$ in the Radiation Zone, while $\rho _\pm$ in the Force-Free Zone might depend on the regularization of (9)  and on the plasma production rate $\Gamma$. Before we give our plausibility argument in favor of this ``Pulsar No-Hair Theorem'', we must note another feature seen in Fig.3.

Three surfaces, the Force-Free Zone boundary, the light cylinder, and the Null Surface ($\rho _+=\rho _-$) intersect at the critical line $r=1$, $|z|\approx 1$. Inside the Force-Free Zone, and just outside it (except at the equator), the charges move along magnetic surfaces $\psi =const$. If the magnetic surface intersects the Force-Free Zone boundary below the critical line (at low altitudes), only positrons flow from the Force-Free Zone into the Radiation Zone. If the magnetic surface intersects the Force-Free Zone boundary above the critical line (at high altitudes), both positrons and electrons flow from the Force-Free Zone into the Radiation Zone. 

This change in the character of the inflow from the Force-Free Zone into the Radiation Zone is expected. Indeed, according to Eq.(8), just outside the Force-Free Zone, where $E_0$ is small, the poloidal velocities are $v_\pm \propto B_tE_p\pm B_0B_p$, where $B_t$ is the toroidal magnetic field, and $E_p$, $B_p$ are the poloidal electric and magnetic fields, $B_0^2=B_t^2+B_p^2-E_p^2$. Then the product of velocities is $v_+ v_- \propto E_p^2-B_p^2$. In the Force-Free Zone, and just outside it, $E_p=rB_p$ (see below), giving $v_+ v_- \propto r^2-1$. Thus, at low altitudes, at $r<1$, only one species can flow into the Radiation Zone, while at high altitudes, at $r>1$, both species can flow into the Radiation Zone.

After these observations, we can attempt a proof that there exists a unique AE flow in the Radiation Zone bounded by an arbitrary (non-AE) Force-Free Zone. The Force-Free Zone is described by the pulsar version of the Grad-Shafranov equation (Scharlemant-Wagoner equation \cite{Scharlemant}):
\be\label{fff}
(1-r^2)\nabla ^2\psi -{2\over r}\partial _r\psi =-A{dA\over d\psi }, 
\ee
with $\phi =\psi$ and $A=A(\psi )$.
Let $\psi _c$ be the magnetic surface containing the critical line, here defined as the intersection of the light cylinder and the boundary of the Force-Free Zone. Inside the critical magnetic surface, for $\psi > \psi _c$, the Scharlemant-Wagoner equation is elliptic, and thus one can find a solution parametrized by three arbitrary functions:
\begin{itemize}
\item the shape of the Force-Free Zone boundary
\item the current $A(\psi )$
\item the value of $\psi$ on the Force-Free Zone boundary
\end{itemize}

Now, ignoring for a while the magnetic surfaces $\psi < \psi _c$, we need to calculate the corresponding AE magnetosphere outside the Force-Free Zone. It is clear that the result must be unique. Indeed, for the AE calculation, we know the shape of the boundary, the boundary $\psi$, meaning the normal magnetic field, the boundary $\phi$, meaning the tangential electric field, and, as only the positrons may enter, we know how many positrons are injected across the boundary from the known $A(\psi )$.

Suppose the AE calculation saturates, giving some $\psi$, $\phi$, and $A$ in the Radiation Zone. Also assume, in agreement with all our figures, that the isolines of these three fields cross the boundary of the Force-Free Zone smoothly (have continuous derivatives; in other words, the non-corotating part of the Force-Free Zone boundary is not a singular charge/current sheet). Then we have {\it two} matching conditions at the Force-Free Zone boundary. The smoothness of the $A$ field follows from the smoothness of the $\psi$ and $\phi$ fields, because the poloidal current flows along the magnetic surfaces both in the force-free field and in the  AE field with infinitesimal $E_0$. We therefore have to satisfy just two boundary conditions by choosing the three above-listed arbitrary functions. This seems to lead to an arbitrariness of the solution. We propose, in agreement with Fig.3, that in the elliptic region, $\psi > \psi _c$, the position of the Force-Free Zone boundary is not arbitrary -- it must coincide with the Null Surface. Then the number of adjustables matches the number of the boundary conditions, which might lead to a unique solution.

It would seem, however, that at high altitudes, for $\psi < \psi _c$, our counting argument breaks down. As we have seen, at high altitudes both electrons and positrons can flow across the boundary of the Force-Free Zone. Then the boundary condition for the AE calculation adds an extra arbitrary function -- the multiplicity of the plasma injected into the Radiation Zone. The plasma multiplicity at injection, together with the three above-listed arbitrary functions, gives four adjustables, while we have only two boundary conditions. We will assume, in approximate agreement with Fig.3, that the plasma multiplicity in the force-free part of the region $\psi < \psi _c$ is in fact zero. Then the plasma multiplicity at injection into the Radiation Zone must be zero, reducing the number of adjustables to three -- still one too many. But, as explained by Contopoulos, Kazanas and Fendt \cite{Contopoulos}, the nature of the Scharlemant-Wagoner equation (12) changes for $\psi < \psi _c$. Eq.(12) provides its own boundary condition on the light cylinder, ${2\over r}\partial _r\psi =A{dA\over d\psi }$; the requirement of smoothly crossing the light cylinder then fixes $A(\psi)$ for $\psi < \psi _c$. This reduces the number of adjustables to two, equal to the number of the boundary conditions, which might lead to a unique solution.

Alternatively, and also not in contradiction with the numerics, one can assume that the high altitudes, $\psi < \psi _c$, are not truly force-free. There is a small $E_0$ in this region, explaining the observed absence of positrons. Then there really exists one more boundary of the Force-Free Zone within the light cylinder, the surface $\psi =\psi _c$ (not shown in the figures, while the shown boundary of the Force-Free Zone is in fact fictitious beyond the light cylinder, and must be removed). The new part of the Force-Free Zone boundary coincides with the surface of zero poloidal current density -- the surface of maximal value of $|A|$, emanating from the star and terminating at the light cylinder. If this alternative is correct, the boundary of the Force-Free Zone consists of the following three special surfaces: (i) the maximal $|A|$ surface, (ii) the Null Surface, (iii) the boundary of the Corotation Zone. 

Needless to say, the above arguments are not a proof.

\section{Conclusion}

The weak pulsar magnetosphere calculated in this paper by a full AE simulation is very close to the phenomenologically successful magnetosphere calculated by a simplified procedure \cite{Gruzinov}. While some details may change, it appears that pulsar gamma-ray emission is basically understood. 

Still, the following calculations seem doable and interesting

\begin{itemize}

\item Using full AE in 3D, one can calculate energy-resolved lightcurves. Then one can measure all parameters (distance, magnetic dipole moment, moment of inertia, spin-dipole angle, observation angle) for all weak Fermi pulsars. 

\item Dying (millisecond) pulsar  must be a perfect target for a PIC simulation (with modeled or exact pair production), because here the screening of the proper electric field is only partial, and may turn out to be numerically treatable.

\item Strong pulsar should be doable by adding relevant plasma production by photon-photon collisions to the AE simulation performed in this paper.

\item To really prove the Weak Pulsar No-Hair Theorem, and get the numerically-exact axisymmetric weak pulsar magnetosphere, one should solve (numerically, of course) the combined Grad-Shafranov-AE problem described at the end of \S IV. Alternatively, one may improve our bad numerics and convincingly demonstrate that the pure-AE calculation does not depend on the regularization and the  plasma production prescriptions.

\end{itemize}

\thanks

I thank the participants of the recent Pulsar workshop at Princeton, organized by Sasha Philippov and Tolya Spitkovsky. It was very encouraging to see how various different approaches close in on the unique pulsar solution. 

\appendix

\section{Details of Numerics}

\begin{enumerate}

\item Grid and fields: square r-z grid; $E_r$, $j_{\pm r}$, $B_z$ are at the r-edges; $E_z$, $j_{\pm z}$, $B_r$ are at the z-edges; $E_\phi $, $\rho _\pm$, $j_{\pm \phi }$ are at the vertices; $B_\phi $ are at the faces. 

\item The star is at $r^2+z^2<r_s^2$, with $r_s=0.25$. Inside the star
\be
\rho _\pm =0,
\ee
\be
{\bf j}=\sigma _s({\bf E}+(\hat{z}\times {\bf r})\times {\bf B})+j_e\hat {\phi },
\ee
where the conductivity of the star is large, $\sigma _s=200$. The external toroidal current $j_e$ is chosen so as to give the magnetic dipole moment $\mu =1$.

\item Diffusive regularization: diffusion $D\nabla ^2\rho _\pm$ is added to eq.(7), with small diffusivity, $D=0.0006$. Doubling and halving the diffusivity changes the results by about 10\% .

\item Velocities regularization. As written, eq.(9) gives
\be
B_0=\sqrt{B^2-E^2+E_0^2}~{\rm sign} ({\bf E}\cdot {\bf B}).
\ee
We replace
\be
{\rm sign} ({\bf E}\cdot {\bf B})\rightarrow {{\bf E}\cdot {\bf B}\over \sqrt{ ({\bf E}\cdot {\bf B})^2+\alpha E^2B^2} },
\ee
with small $\alpha = 0.001$. Doubling and halving $\alpha$ does not noticeably change the results.

\item Plasma production. We use constant plasma production rate $\Gamma$ operating in a layer $1.1r_s^2<r^2+z^2<1.5r_s^2$, provided the altitude is above some cutoff, or if the proper electric field is large, $E_0^2>\alpha E^2$, with the same small $\alpha$ as above, just to avoid introducing new parameters. The plasma production rate $\Gamma$, and the altitude cutoff are adjusted so as to minimize the volume averaged $E_0^2$ in the entire layer $1.1r_s^2<r^2+z^2<1.5r_s^2$. 

\item Interpolation. The non-diffusive part of the fluxes ${\bf j}_\pm$, namely $\rho _\pm {\bf v}_\pm$, is calculated as follows. For each vertex, we calculate the interpolated electric and magnetic field, and use the regularized eq.(8) to calculate the velocity components. If, say, $v_{+r}>0$, we add the current $\rho _+v_{+r}$ to the component $j_{+r}$ on the edge which is to the right of the vertex, etc.

\item Initial condition: everything is zero.

\item Boundary conditions: no boundary conditions are needed at the surface of the star. The outer boundary conditions are outgoing for the electromagnetic fields and absorbing for the densities. 

\item Resolution. Full in $r$ and half in $z$ part of the simulation boxes are shown. The small boxes are (200,400), the large boxes are (600,1200).

We must call our numerics dubious for two reasons. This author can't be impartial and after a year of trial and error one can get many things numerically. The results do show some dependence on the regularization and plasma production prescriptions.

\end{enumerate}

\end{document}